\documentclass[prb,twocolumn]{revtex4}
\usepackage{graphicx}
\usepackage{dcolumn}
\usepackage{amsmath}

\makeatother

\begin{document}

\title{Hourglass of constant weight}

\author{Volker Becker and Thorsten P\"oschel}
\affiliation{Charit\'e, Augustenburger Platz 1, 13353 Berlin, Germany}

\date{\today}

\begin{abstract}
In contrast to a still common believe, a steadily flowing hourglass changes its weight in the course of time \cite{ShenScott:1985}. We will show that, nevertheless, it is possible to construct hourglasses that do not change their weight.
\end{abstract}

\maketitle

In the state of steady flow, the center of mass of the sand in an hourglass is permanently moving downward. The absolute velocity of the center of mass decreases leading to an upward acceleration of the center of mass. This simple argument was worked out by Shen and Scott \cite{ShenScott:1985} and led to the conclusion that a suspended at a balance running hourglass appears to be of larger weight than the pure weight of the hourglass and the contained sand, that is, a flowing hourglass appears to be heavier than a non-flowing one by a certain quantity $\Delta m$. While the time dependence of $\Delta m(t)$ depends on the specific construction of the hourglass, it was shown\cite{ShenScott:1985}  that $\Delta m(t)$ is always a positive function that decreases with time. 

\begin{figure}[b]
  \centerline{\includegraphics[width=3.5cm,clip]{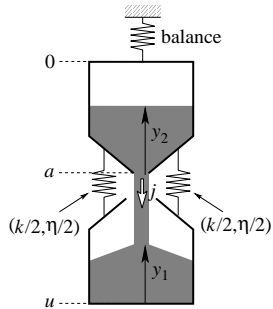}}
  \caption{Hourglass in steady-state motion. The lower container is suspended at the upper one by damped springs. Choosing appropriate parameters of the spring constants the weight of the hourglass is independent of time.}
  \label{fig:hourglass}
\end{figure}

The finding by Shen and Scott naturally provokes the question whether we can invent an hourglass whose total weight in the steady flowing state is independent of time. It turns out that such a construction is possible. Consider the hourglass shown in Fig. \ref{fig:hourglass}.

The vertical component of the center of mass of the hourglass, $y_{\rm cm}$, is described by 
\begin{multline}
  My_\mathrm{cm} = 
  \int\limits_{u}^{u+y_1}\mathrm{d} y\, \rho y A_1(y-u) + 
  \int\limits_{a}^{a+y_2}\mathrm{d} y\, \rho y A_2(y-u)\\ +
  \tilde{m}_1 y_\mathrm{cm}^{(1)} + 
  \tilde{m}_2 y_\mathrm{cm}^{(2)} 
\,,
\end{multline}
where $A_1(y)$ and $A_2(y)$ are the cross-sections of the lower and upper container, $\tilde{m}_1$ and $\tilde{m}_2$ are the masses of the (empty) containers, $y_\mathrm{cm}^{(1)}$ and $y_\mathrm{cm}^{(2)}$ are the vertical component of their center of masses, $\rho$ is the mass density of the sand, and $M$ is the total mass of the hourglass and the contained sand. For the definition of $y_1$, $y_2$, $a$, and $u$ see Fig. \ref{fig:hourglass}. The velocity of the center of mass reads
\begin{multline}
  M\dot{y}_\mathrm{cm} = 
  \rho\left(\dot{u}+\dot{y}_1\right)\left(u+y_1\right)A_1\left(y_1\right)
  -\rho\dot{u} u A_1(0) \\
  + \rho \dot{y}_2\left(a+y_2\right)A_2\left(y_2\right)
  + \tilde{m}_1\dot{u}
  + \rho\dot{u}\!\!\!\int\limits_{u}^{u+y_1}\!\!\!\mathrm{d}y y \left.\frac{\mathrm{d}A_1}{\mathrm{d}u}\right|_{(u-y_1)}
\end{multline}
Using $\left.\mathrm{d}A_1/\mathrm{d}u\right|_{(y-u)}=-\left.\mathrm{d}A_1/\mathrm{d}y\right|_{(y-u)}$ the last term reads $m_1(t)\dot{u}+\rho\dot{u}uA_1(0)-\rho\dot{u}\left(u+y_1\right)A_1\left(y_1\right)$, where $m_1(t)$ is the mass of the sand in the lower compartment. With the definition of the mass flow 
\begin{equation}
  j=\rho A_1\left(y_1\right) \dot{y}_1 = -\rho A_2\left(y_2\right)\dot{y}_2
\end{equation}
we obtain 
\begin{equation}
M\dot{y}_\mathrm{cm} = \left[m_1(t) + \tilde{m}_1\right] \dot{u} + j \left(u-a+y_1-y_2\right)
\end{equation}
and, thus,
\begin{equation}
M\ddot{y}_\mathrm{cm} = \left[m_1(t)+\tilde{m}_1\right] \ddot{u} + 2 j \dot{u} + \frac{j^2}{\rho}\left[\frac{1}{A_1\left(y_1\right)} + \frac{1}{A_2\left(y_2\right)}\right]\,.
\label{eq:ytd2}
\end{equation}
For the case of constant cross-section, as sketched in Fig. \ref{fig:hourglass}, $A_1\left(y_1\right)=A_2\left(y_2\right)=A$ and with $m_1(t)=j\,t$, the change of weight of the hourglass as measured by the balance is
\begin{equation}
  M\ddot{y}_\mathrm{cm} = \left[jt+\tilde{m}_1\right] \ddot{u} + 2 j \dot{u} + \frac{2j^2}{\rho A}\,.
\end{equation}
This equation provides a condition for the time-dependent position $u(t)$ of the lower compartment to keep the weight of the hourglass constant, $M\ddot{y}_\mathrm{cm}=0$. A particular solution of the latter equation is 
\begin{equation}
\dot{u}=-\frac{j}{\rho A}\,,
\label{eq:udot}
\end{equation}
that is, the lower container moves downward at constant velocity. 

Assume the lower container is suspended via damped linear springs as shown in Fig. \ref{fig:hourglass}. Let us see whether this system fulfills the condition Eq. \eqref{eq:udot} as a {\em stable} solution. Newton's equation of motion for the position of the container $u(t)$ reads
\begin{equation}
\left[m_1(t)+\tilde{m}_1\right]\ddot{u} + \eta\dot{u}+ku = -m_1(t)g- F_j - \tilde{m}_1 g\,,
\end{equation}
where $k/2$ and $\eta/2$ are the elastic and dissipative constants of the springs and $F_j$ is the force that originates from the impact of the free falling particles at the surface according to the constant flow rate $j$. Using the previous expressions for $m_1(t)$ and abbreviating the last two terms by the constant $K$ we obtain
\begin{equation}
\left(jt+\tilde{m}_1\right)\ddot{u}+\eta\dot{u}+ku=-jtg + K
\end{equation}
with the solution
\begin{multline}
  u(t)=\left(jt+\tilde{m}_1\right)^\frac{j-\eta}{2j}\left[C_1J_{1-\frac{\eta}{j}}\left(2\sqrt{\frac{jkt+k\tilde{m}_1}{j^2}}\right) \right.\\ 
\left. + C_2 Y_{1-\frac{\eta}{j}}\left(2\sqrt{\frac{jkt+k\tilde{m}_1}{j^2}}\right)\right]
-\frac{jtg+K}{k} + \frac{\eta j g}{k^2}\,,
\label{eq:usol}
\end{multline}
where $J_n(x)$ is the Bessel function of order $n$ and $Y_n(x)$ is corresponding Weber function and $C_1$ and $C_2$ are integration constants. For large $t$ these functions behave as\cite{AbramowitzStegun} 
\begin{equation}
\begin{split}
J_n(x)& \simeq \sqrt{\frac{2}{\pi x}}\left[\cos\left(x-\frac{n\pi}{2} - \frac{\pi}{4}\right)\right]\\
Y_n(x)& \simeq \sqrt{\frac{2}{\pi x}}\left[\sin\left(x-\frac{n\pi}{2} - \frac{\pi}{4}\right)\right]\,,
\end{split}
\end{equation}
that is, the absolute value of the term in square brackets in Eq. \eqref{eq:usol} decays as $t^{-1/4}$. Therefore, for stability of the solution, the prefactor of this term must not  increase steeper than $t^{1/4}$, i.e., $\eta/j > 1/2$. With this condition, we obtain from Eq. \eqref{eq:usol} for the asymptotic velocity $\dot{u}_\infty = jg/k$. Comparing with Eq. \eqref{eq:udot} we obtain $k=\rho A g$.

In conclusion, we find that although a simple hourglass in the state of steady flow changes its weight in  the course of time\cite{ShenScott:1985}, it is possible to construct hourglasses whose weight remains constant. To this end, as sketched in Fig. \ref{fig:hourglass}, the lower container should be suspended by damped linear elastic springs of total force \begin{equation}
F\left(u,\dot{u}\right)=-ku-\eta\dot{u}
\end{equation}
with elastic and dissipative constants
\begin{equation}
  \begin{split}
    k &=\rho A g\\
    \eta &>j/2\,.
  \end{split}
\end{equation}

\end{document}